Routledge
Taylor & Francis Group

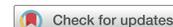

# Basic thermodynamics to save lives

Mouhamadou Thiam

Département d'enseignement du secondaire et des ressources humaines, Université de Moncton, Moncton, New Brunswick, Canada

**ABSTRACT**
Students have several misconceptions about thermodynamics. However, an understanding of basic concepts in this branch of physics remains fundamental to explain thermal phenomena. The purpose of this activity was to extend students' conceptual understanding of heat, temperature, energy, and thermal insulators and conductors. Two groups of students in grade seven created insulation boxes to keep sheep organs in a hypothermic state for the longest time possible. The activity incorporated connections to real life, autonomy, and scientific investigations to motivate the students. The data indicated that the activity was sufficiently challenging for the students and that they were able to apply their understanding of both the corresponding concepts and scientific and engineering practices.

## Introduction

Students often come to class with misconceptions about the world around them (Duit 2009; Thouin 2004). Misconceptions represent understandings, conceptions or beliefs held by students, and are different from those of scientists (Murphy and Alexander 2008), so they form barriers to learning science (Guzzetti et al. 1993). Although fundamental to most physical sciences, thermodynamics is difficult to learn. In fact, students frequently have robust misconceptions about concepts of heat and temperature (Reiner et al. 2000). For instance, Lewis and Linn (1994) found several intuitive conceptions about thermal phenomena shared among middle school students. The most unexpected ones were those of metals as insulators (e.g., "metals absorb, hold, or attract cold in cool environments"); insulators as conductors (e.g., "insulators conduct heat fast and heat leaves so insulators don't feel hot"); and hot wool (e.g., "wool warms things up"). Holding such intuitive conceptions have consequences, evident through their surprising extensions (Lewis and Linn 1994). For example, the extension made by students who state "wool warms things up" is that "wool cannot be used as an insulator for cold objects since it warms things up". For Wiser and Amin (2001), a significant barrier to learning thermodynamics is that students conceive heat as hotness because they do not differentiate it from temperature; they are resistant to the fact that heat is proportional to the amount of material. From a scientific point of view, heat is a spontaneous transfer of energy between two substances at different temperatures. Energy is present in a substance in the form of particle agitation and is called internal kinetic energy. Temperature measures the average internal kinetic energy of particles in a substance. Therefore, a conceptual understanding of heat, temperature, and energy remains fundamental to explain thermal phenomena.

To achieve this understanding, the teacher (author) developed and implemented an activity in which students made an insulation box to keep sheep organs in a hypothermic state. The activity is that of the Elaboration Phase based on the BSCS 5E Instructional Model (Bybee et al. 2006). Thus, the goal was to extend students' conceptual understanding of heat, temperature, energy, and thermal insulators and conductors by giving them the opportunity to apply what they have learned. As presented in Table 1, the activity seems to facilitate the integration of science and engineering practices, crosscutting concepts, and disciplinary core ideas (National Research Council 2012).

Table 1. Alignment to the Next Generation Science Standards Lead States (2013).

| Performance expectation | |
| --- | --- |
| Students who demonstrate understanding can: | |
| MS-PS3-3. | |
| Apply scientific principles to design, construct, and test a device that either minimizes or maximizes thermal energy transfer. | |
| Science and engineering practices | Constructing Explanations and Designing Solutions |
| Disciplinary core ideas | Constructing explanations and designing solutions in 6–8 builds on K–5 experiences and progresses to include constructing explanations and designing solutions supported by multiple sources of evidence consistent with scientific ideas, principles, and theories. |
| | • Apply scientific ideas or principles to design, construct, and test a design of an object, tool, process or system. |
| | PS3.A: Definitions of Energy |
| | • Temperature is a measure of the average kinetic energy of particles of matter. The relationship between the temperature and the total energy of a system depends on the types, states, and amounts of matter present. |
| | PS3.B: Conservation of Energy and Energy Transfer |
| | • Energy is spontaneously transferred out of hotter regions or objects and into colder ones. |
| | ETS1.A: Defining and Delimiting an Engineering Problem |
| | • The more precisely a design task's criteria and constraints can be defined, the more likely it is that the designed solution will be successful. Specification of constraints includes consideration of scientific principles and other relevant knowledge that is likely to limit possible solutions. (secondary) |
| | ETS1.B: Developing Possible Solutions |
| | • A solution needs to be tested, and then modified on the basis of the test results in order to improve it. There are systematic processes for evaluating solutions with respect to how well they meet criteria and constraints of a problem. (secondary) |
| Crosscutting concept | Energy and Matter |
| | The transfer of energy can be tracked as energy flows through a designed or natural system. |

Throughout the activity, students were successfully engaged in potentially motivating tasks with live materials. For instance, the teacher tried to generate situational interest by linking the course with real life through the topic of organ donation and a presentation given by a researcher of organ preservation.

## Participants

The participants were 42 (19 boys and 23 girls) grade seven students (12- to 13-year-olds). The students came from a predominantly English-speaking city in southeastern New Brunswick, Canada. These adolescents enrolled in the Early French Immersion program at a K-8 public school. Thus, commencing in grade three, they learned French by attending the majority of their courses such as science, in that language. The teacher taught the two intact groups of 21 students for an equal amount of time, each class in the science laboratory. The classes consisted of three one-hour periods per week. The activity required eighteen one-hour periods and occurred at the end of a unit on conservation of energy and energy transfer.

Prior to engaging in the learning activity, the teacher formally introduced the concepts of heat, temperature, energy, and thermal conductors and insulators after students had completed several investigations (see Appendix 1) adapted from the Thermal Concept Evaluation (TCE; Yeo and Zadnik 2001). The advantage of the TCE is that the general question presents a scenario that reflects a real-life situation about thermal physics; it is followed by statements of scientific and common students' alternative conceptions. The purpose of these investigations was to get the students to explore the scientific concepts under study through experimentation. Thus, the students had to ask testable questions, formulate hypotheses, identify and isolate variables, propose procedures, collect data, and analyze and organize the results.

## Engaging students (two periods)

Organ donation has become a global issue as there are not enough donors for the number of possible recipients. Thus, the consequences of this organ shortage range from the preventable deaths of patients to the emergence of illegal organ transplantation markets. An indirect way to raise students' awareness about organ donation through science is to allow them to design, build, and test an insulation box that keeps sheep organs in a hypothermic state for the longest time possible. In fact, placing an organ in a hypothermic state helps to prevent tissue damage by slowing cellular metabolism (Guibert et al. 2011).

First, in order to trigger their interest in the topic, students watched a documentary about families involved in the process of organ donation (Thompson 2016). Many seemed sad after watching the documentary. The students asked several questions about why some people do not register as donors. This led students to identify a number of personal and religious beliefs. Afterwards, the teacher informed students that they would be focusing on organ conservation in Science, and that they would be exploring some other related issues in classes such as Language Arts, Mathematics, and Health.

After the discussion about organ donation, students discovered the performance task presented in Figure 1. The teacher informed them they must apply the previously learned concepts of heat, temperature, energy, and thermal insulators and conductors to design and explain how their insulation box worked.

Students then received the rubric, presented in Table 2, and an explanation of how the teacher would use it to assess their engagement in engineering and science practices. The rubric focuses on the engineering practice of designing solutions and the science practices of: planning and investigations, analyzing and interpreting data, and constructing explanations. McNeill, Katsh-Singer, and Pelletier (2015) developed the indicators of performance to assess the science practices, while the teacher created the ones for the engineering practice to assess the insulation boxes. The rubric of McNeill, Katsh-Singer, and Pelletier was chosen because it provides simple descriptions of the science practices that middle school students can understand easily. For clarification, the teacher answered students' various questions, such as: the length of time the insulation box will keep the organ in a hypothermic state, the number of times the data they collected will need to be replicated and the type of materials they will use.

Finally, students formed cooperative groups by self-selecting 3–4 peers. Relevant social skills to use when working in collaboration with others were discussed. For instance, students had the responsibility to ensure that everyone on their team performed the various tasks assigned to them. It was instructed that when internal conflicts arise, students must inform the teacher so they can find solutions. Throughout the activity, the teacher was actively engaged, supervising, observing, and questioning the students about what they were doing. When a student seemed distracted from the task, they engaged in a discussion to understand the reasons and they looked for solutions.

## Designing and building insulation boxes (nine periods)

The teacher informed students that the materials and tools presented in Table 3 would be made available to them. This choice of materials was motivated by the fact that they were most

---

In 2016, 4500 people were on a waiting list for an organ transplant (Government of Canada 2019). In the same year, 2835 transplants were performed, but 260 people died due to a lack of organ donors.

When a person who is on a donor list dies, up to 8 organs can be taken. Then, they are transported to a hospital where recipients are located. During transport, the organs must be preserved in a hypothermic state: between 4 °C and 8 °C.

To simulate the transport of an organ from a donor to a recipient, you will create a box that will keep a sheep organ in a hypothermic state for the longest time possible. You will need to plan and carry out investigations to collect data, analyze the data to create a representation, and use the concepts to explain how your box works.

**Figure 1.** Performance task.

**Table 2.** Rubric.

| Science and engineering practices | Level 1 (Not present) | Level 2 Emergent | Level 3 Proficient | Level 4 Exemplary |
| --- | --- | --- | --- | --- |
| Planning and carrying out investigations | Students do not design or conduct investigations. | Students conduct investigations, but these opportunities are typically teacher driven. Students do not make decisions about experimental variables or investigational methods (e.g., number of trials). | Students design or conduct investigations to gather data. Students make decisions about experimental variables, controls, or investigational methods (e.g., number of trials). | Students design and conduct investigations to gather data. Students make decisions about experimental variables, controls, and investigational methods (e.g., number of trials). |
| Analyzing and interpreting data | Students may record data but do not analyze data. | Students work with data to organize or group the data in a table or graph. However, students do not recognize patterns or relationships in the natural world. | Students work with data to organize or group the data in a table or graph. Students make sense of data by recognizing patterns or relationships in the natural world. | Students make decisions about how to analyze data (e.g., table or graph) and work with the data to create the representation. Students make sense of data by recognizing patterns or relationships in the natural world. |
| Constructing explanations | Students do not create scientific explanations. | Students attempt to create scientific explanations, but students' explanations are descriptive instead of explaining how or why a phenomenon occurs. Students do not use appropriate evidence to support their explanations. | Students construct explanations that focus on explaining how or why a phenomenon occurs. Students do not use appropriate evidence to support their explanations. | Students construct explanations that focus on explaining how or why a phenomenon occurs and use appropriate evidence to support their explanations. |
| Designing solutions | Students do not create a box that can keep the sheep organ between 4 °C and 8 °C. | Students create a box that can keep the sheep organ between 4 °C and 8 °C for at least 30 minutes. Students cannot replicate these results. | Students create a box that can keep the sheep organ between 4 °C and 8 °C for at least 30 minutes. Students can replicate these results at least once. | Students create a box that can keep the sheep organ between 4 °C and 8 °C for at least 30 minutes. Students can replicate these results at least three times. |

commonly used by students over the last two years refining this activity. However, students could work with other materials if they wished, but they were responsible for providing them. The sheep organs were selected to make the project as lifelike as possible. These organs were individually sealed in plastic bags and randomly assigned to groups.

Students sketched a design for their insulation box. They used the internet to research the thermal properties of the materials they might use to construct the box. They noted the reasons for choosing each material and the source of information. The teacher made sure to remind each group of the importance of relating the materials with the concepts of heat, temperature, energy, and thermal insulators and conductors that were learned in class. For instance, a group that sketched an insulation box using cardboard, polystyrene, plastic container and water wrote:

> The thermal properties of cardboard make it a good insulator, as it is a relatively poor conductor of heat… Expanded polystyrene consists of trapped air bubbles that prevent heat from being transferred, which makes polystyrene an excellent insulator… We use a plastic container because we don't want water leaking… Water has a conductivity of 0.606.[1]

Subsequently, students presented their sketches in front of the class using a multimedia projector and explained the reasons for their choice of materials. The presentation to the class was very important as groups related the functioning of their insulation boxes to the concepts of heat, temperature, energy, and thermal insulators and conductors. Students made arguments such as the one presented in the previous paragraph that supported how energy transfer could be optimized to keep the sheep organ in a hypothermic state. Afterwards, the presenters answered their classmates' questions. When unable to give an answer, someone in the class attempted to provide one. Also, the classmates provided relevant feedback. For example, one group planned to place a small fan inside their box to lower the temperature. Other students advised them to add ice packs to the box as a fan would only circulate the air and not lower the temperature. After the question-and-answer and comments session, the teacher asked the presenters what changes they were going to make to optimize their design. Students made changes on paper to their designs. Figure 2 shows the final sketch for the insulation box of one group. The peers approved the final designs. This teaching strategy of involving students in the decision-making supports autonomy which enhances engagement (Ames 1990, 1992; Maehr and Midgley 1991), and engagement is presumed to positively influence achievement (Fredricks, Blumenfeld, and Paris 2004).

Table 3. Approximate quantities, and specifications of materials and tools for one class.

| Materials and tools | Quantity | Specifications |
| --- | --- | --- |
| Sheep organs | 8 | Preserved brains, eyes, hearts and kidneys |
| Polystyrene | 4 | $1.22 \times 2.44 \times .03$ m |
| Cardboard | 10 | $30.48 \times 30.48 \times 15.24$ cm |
| Saran Wrap | 2 | $.3 \times 30$ m |
| Aluminum foil | 2 | $.3 \times 8.2$ m |
| Masking tape | 4 | $.03 \times 54.86$ m |
| Ice packs | 50 | 236.59 mL |
| Utility knives | 10 | Retractable; automatic locking |
| Hot glue stick | 85 | $1.12 \times 25.4$ cm |
| Glue guns | 10 | Low temperature |
| Goggles | 22 | Coverall-style |
| Thermometers | 10 | Length: 30.5 cm; Range: $-20\,°C$ to $110\,°C$ |

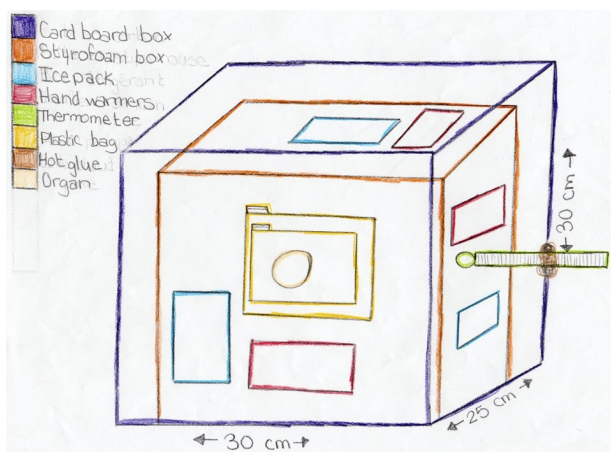

Figure 2. Sketch of an insulation box made by students.

Finally, each group was assigned a cupboard in the laboratory in which they stored their materials. Laboratory safety rules were reviewed and applied. Students had to wear goggles during construction and testing and were instructed to wash their hands thoroughly after each class. After the teacher demonstrated the proper use and storage of the utility knives, students cut materials in a specific part of the room and did not cut toward their bodies or toward other individuals. Figure 3 shows some actual insulation boxes and students in the process of constructing those boxes while applying some safety rules such as wearing goggles.

## Planning and carrying out investigations (five periods)

After creating their insulation boxes, students planned an investigation to collect data. They wrote a testable question, formulated hypotheses, identified variables (dependent, independent and controlled), listed materials, and proposed a procedure. One group wrote:

> Testable question—"Does the amount of ice packs affect the temperature inside of the box over a period of time?"; hypothesis—"We believe that the number of ice packs will affect the temperature inside of the box for a period of time because the air inside of the box will transfer thermal energy to the ice pack. Thus, the more ice packs you have, the lower the temperature will be inside of the box for a longer period of time."; independent variable—"amount of ice"; dependent variable—"temperature inside of the box"; and controlled variables—"box, place to carry out the tests, type of ice pack, organ, thermometer, initial temperature."

The variable manipulated by most of the groups was the number of ice packs inside their insulation boxes. Depending on the time of year that this project is launched, it will be necessary to engage students in planning and carrying out investigations. As a result, this activity will require more time. The rubric in Appendix 2 (Government of New Brunswick 2017) gives teachers ideas for what is truly important in this activity. Some students intended to collect their data every hour or once a day. To guide them in choosing more realistic time intervals, students were asked questions such as: *how will you know if the temperature has changed within a wide time interval?* They were also reminded that the exact goal for the box is to keep the sheep organ between 4 °C and 8 °C for as long as possible.

Following the planning of their investigation, students carried out their procedures. To begin,

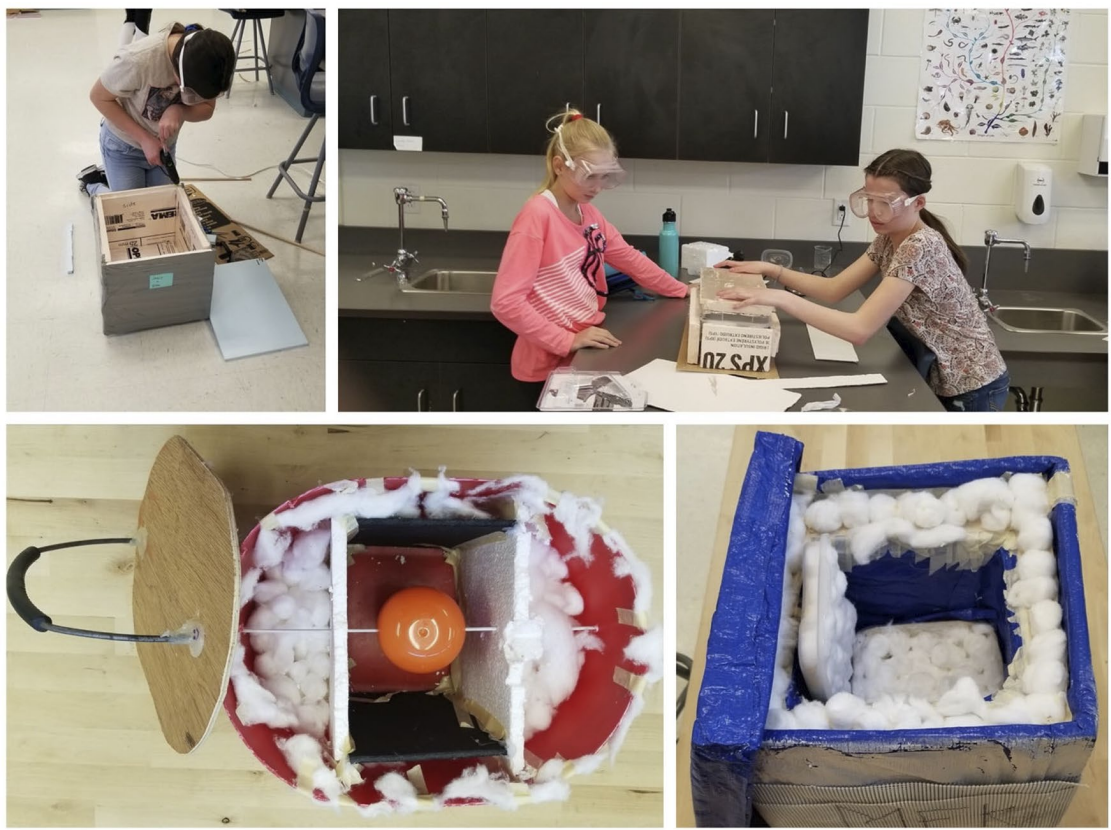

**Figure 3.** Students at the task and prototypes of insulation boxes.

the sheep organs and the ice packs were taken out of the refrigerator respectively, at 4 °C and out of the freezer at −18 °C. At the end of each period, students placed them in separate storage containers that the teacher put back in the refrigerator and freezer. For safety concerns, the teacher instructed students not to tear the bags containing the sheep organ. Students measured the temperature inside the insulation box as a function of time because thermal equilibrium would be reached inside after a certain time. Therefore, it is assumed that the temperature of the sheep organ corresponds to the air temperature within the insulation box measured by students. The latter used an iterative approach by making changes based on the data. The investigation ran until the temperature inside the insulation box remained constant for approximately five minutes. If this temperature was outside of the range of 4 °C to 8 °C, students ended their investigation and made some changes. For example, some made changes to their insulation boxes, while others simply added or removed ice packs before starting their investigation again. Otherwise, students continued to collect their data until the end of the period or until the temperature fell outside of the desired range in less than 30 minutes. After more than a dozen trials, several groups were able to keep the organs in a hypothermic state.

Finally, a doctor-researcher from a local university, whose specialization is organ conservation and transportation, was invited to the school to give a presentation on his research. Some students had the opportunity to present their insulation boxes and their preliminary results. The purpose of inviting this researcher was to maintain students' interest in their activity. The presence of an expert can allow students to make connections between the activity materials and real life (Linnenbrink-Garcia, Patall, and Messersmith 2013; Thiam and Lirette-Pitre 2020).

### Analyzing and interpreting data, and constructing explanations (two periods)

Students applied the concepts of heat, temperature, energy, and thermal insulators and conductors to their data sets and graphs to explain how their insulation box functioned. First, they organized their data in a table and drew line graphs as shown in Figure 4. In addition, groups who successfully replicated their data calculated the average temperature as a function of time, which they represented in their graph.

Then, students interpreted their data by writing the trends in their graphs. For example, one group reported:

> For our test with 17 ice packs, the initial temperature was 21 °C. It dropped to 8 °C after 14 minutes and remained constant for 8 minutes. The temperature dropped to 7.5 °C and remained constant for 14 minutes. At 22 minutes, the temperature dropped to 7.5 °C and remained constant for 14 minutes.

Finally, using the concepts of heat, temperature, energy, and thermal insulators and conductors and incorporating the research they did on different types of materials, students tried to explain why they did or did not manage to keep the organ in a hypothermic state. For example, this group successfully applied the concepts of heat, energy and temperature:

> The average temperature dropped to 14 °C after 22 minutes for the tests with four ice packs. This happened because the temperature of the air in the box was higher than the temperature of the ice packs. The air gave energy to the ice packs.

The application of the concepts of thermal conductors and insulators can also be observed in the reports: "Styrofoam has millions of tiny air bubbles that help slow the progression of heat, as air is a poor conductor of heat."

Along with the concepts focused on during the project, some groups also communicated other discoveries; these students decided to place a fan inside their box to help lower the temperature: "We thought that the fan would have made a difference to the temperature; however, the temperature did not change because the fan does not cool things down, it only moves the air."

In addition to human errors, groups that did not manage to keep the organ in a hypothermic state, despite using thermal insulators, proposed several explanations such as: lack of air tightness, thinness of the box, inconsistent initial temperature of the ice packs between tests, and size of the box. For example, one group mentioned in their report that: "We think our results would have been better if our box had been smaller, because the air in the box would have taken longer to lose energy."

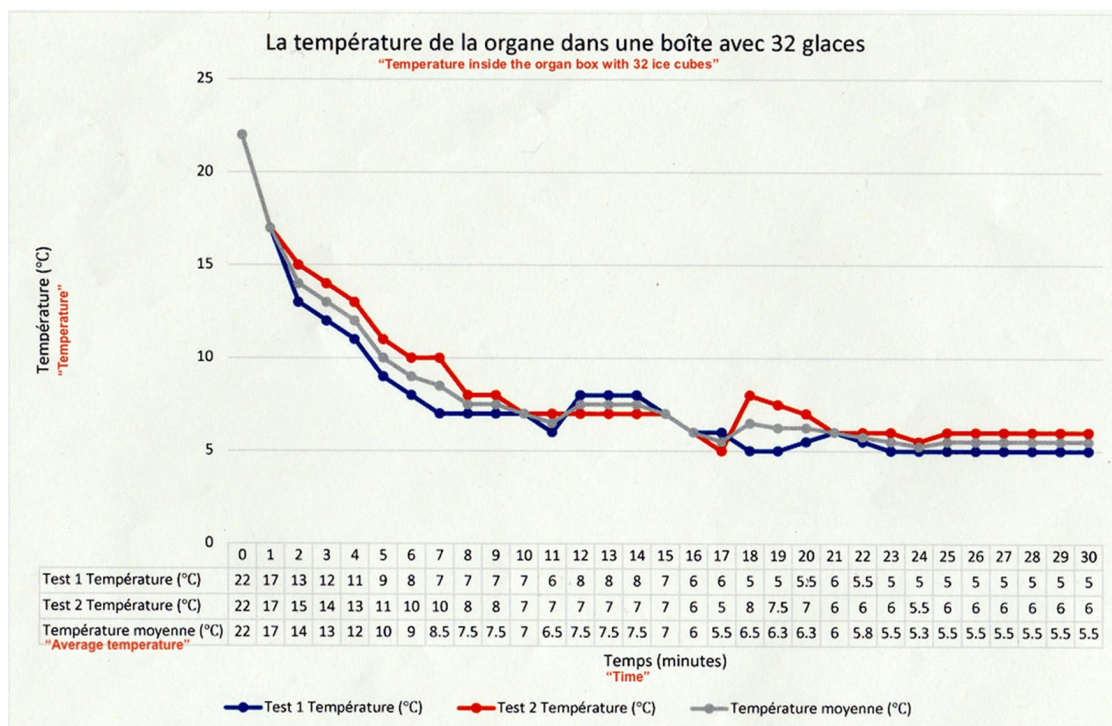

**Figure 4.** Data table and line graphs created by students.

These explanations provided students the basis for suggestions regarding future modifications to their insulation boxes.

### Evaluation

The assessment of the activity was summative, and it targeted four science and engineering practices: planning and carrying out investigations, analyzing and interpreting data, constructing explanations, and designing solutions. The teacher wrote notes on students' observations and conversations as students progressed through the activity. These notes mainly described the help offered to students in regard to planning and carrying out investigations and analyzing and interpreting data. For example, the teacher noted if he guided the students to choose variables or if the students chose the experimental variables. On the other hand, the teacher used the rubric presented in Table 2 to assess students' final reports. Subsequently, all this information was used to determine the level reached by the students. For instance, some students only reached level three in analyzing and interpreting data; the teacher's notes from observations and conversations with the students showed that they decided how to analyze the data, but they did not recognize patterns in their report. Finally, constructing explanations and designing solutions were assessed based exclusively on the information presented in students' reports.

In all practices, most students demonstrated, at a minimum, proficiency. They were able to apply the concepts of heat, temperature, energy, and thermal insulators and conductors to their insulation box. While most students were able to replicate their data, only some groups could do it a minimum of three times. The most efficient insulation box kept the organ in a hypothermic state for three hours. Since the science class only lasted an hour, this group had to carry their box from one class to another.

### Conclusion

The purpose of this activity was to extend students' understanding of basic thermodynamic concepts: heat, temperature, energy, and thermal insulators and conductors. To achieve this goal, the teacher directed students to apply their understanding of these concepts to design, build and test an insulation box that—for most of them—kept a sheep organ in a hypothermic state. Students used different numbers of ice packs to lower the temperature inside their boxes and employed thermal insulators to minimize the transfer of energy from the outside environment to the inside of the box. This activity gave them the opportunity to develop a model to show how their box minimizes heat transfer. Also, students analyzed and interpreted the data on temperature variations collected inside their insulation box to determine the best characteristics to keep the sheep organ in a hypothermic state for the longest time possible.

This project provides the opportunity to integrate other disciplines around the topic of organ donation. All students in grade seven in the same school began this project in Language Arts. They read newspaper articles about organ donation and wrote informative texts on the topic. In Mathematics, they surveyed their family members in order to estimate the number of people registered as donors in the community. In Health, an actor in the community, who had received lungs and a heart via organ donation, gave a presentation to the students. Further, some students created campaigns to raise awareness about organ and tissue donation in the community, such as a 1-K/5-K walk/run.

It is clear that this activity was sufficiently challenging for students and that they were able to apply their understanding of basic thermodynamics concepts and scientific and engineering practices. However, the teacher continues to seek ways to improve it. Firstly, for further in-depth understanding, students could generate similarities and differences between at least two insulation boxes. Based on statistical analysis, students can subsequently suggest improvements to these boxes. Also, they can express their thinking on the concepts by creating conceptual maps. Secondly, this activity is relatively expensive: approximately $200 for a class of 30 students. To reduce the cost of materials, model organs instead of preserved sheep organs are cheaper, in the long run, to use. Further, it is suggested that collecting several cardboard boxes and soliciting

parents to donate some materials would further decrease costs. Finally, it is suggested that teachers use probes that record temperature as a function of time. The probes will encourage students to continue to collect data; additionally, students will not have to carry their boxes from one class to another at the end of the period.

## Acknowledgment

Thanks to Fondation Baxter & Alma Ricard for financial support.

## Note

1. All participants' quotations are translated from French.